\documentclass[journal]{IEEEtran} 
\usepackage{ifpdf}
\usepackage{cite}
\usepackage{amsmath}
\usepackage{array}
\usepackage{float}
\usepackage{subfig}
\usepackage{cite}
\usepackage{graphicx}
\usepackage{amsmath}
\usepackage{epsfig}
\usepackage{epstopdf}
\usepackage{algorithm}
\usepackage{algorithmicx}
\usepackage{algpseudocode}
\usepackage{blkarray}
\usepackage[long]{optidef}
\usepackage[table,xcdraw]{xcolor}
\usepackage{float}
\usepackage{amsfonts}
\usepackage{mathrsfs}

\let\svmaketitle\maketitle
\def\maketitle{\svmaketitle\thispagestyle{empty}}

\ifCLASSINFOpdf
\else
\fi

\begin{document}
\title{Energy Efficient Pairing and Power Optimization for NOMA UAV Network under QoS Constraints}

\author{Irfan Azam,~\IEEEmembership{Student Member, IEEE},~Muhammad Basit Shahab,~\IEEEmembership{Member, IEEE}, \\~and Soo Young Shin,~\IEEEmembership{Senior Member, IEEE}
\thanks{This work is supported by the National Research Foundation of Korea (NRF) grant funded by the Korea government (MSIT) (No. 2019R1A2C1089542).}
\thanks{Irfan Azam and Soo Young Shin are with WENS Lab, Department of IT Convergence Engineering, Kumoh National Institute of Technology, Republic of Korea, (Email: irfanazam@kumoh.ac.kr, wdragon@kumoh.ac.kr)}
\thanks{Muhammad Basit Shahab is with the School of Electrical Engineering and Computing, University of Newcastle, Callaghan, NSW 2308, Australia, (Email: basit.shahab@newcastle.edu.au)}

}

\markboth{IEEE JOURNAL ON SELECTED AREAS IN COMMUNICATIONS}%
{IEEE \MakeLowercase{\textit{et al.}}: Bare Demo of IEEEtran.cls for IEEE Journals}

\maketitle
\begin{abstract}
Due to an increasing number of unmanned aerial vehicles (UAVs) with generally limited battery power, energy efficient data transmission schemes with massive connectivity capabilities are required for future wireless networks. Non-orthogonal multiple access (NOMA) is one of the promising techniques that provide such massive connectivity by allowing superposed data transmission of multiple users over the same resource block. Unlike existing literature, this paper presents a user pairing and power allocation technique for energy efficient and quality of service (QoS) aware NOMA transmission for cellular-connected mobile UAVs. The aim is to minimize the power consumption of the mobile UAVs during uplink data transmission and guarantee their required transmission rate by jointly optimizing the UAV pairing and power allocation. Furthermore, the performance and pairing complexity for mobile UAVs is analyzed to show that the proposed pairing technique efficiently fulfills their QoS requirements. Numerical and simulation results show the performance gain of the proposed technique compared to the conventional NOMA techniques.

\end{abstract}
\IEEEpeerreviewmaketitle
\section{Introduction}
\IEEEPARstart{U}{nmanned} aerial vehicles (UAVs) are gaining popularity with their rapidly growing number and enormous benefits in fifth generation (5G) and beyond 5G (B5G) wireless networks. UAVs are required in many civil, commercial, and military applications such as remote surveillance in hostile territories, search and rescue, traffic control, and photography, etc. The main objective of UAVs involves the collection of data and sending it to the data collection unit (DCU) or the base station (BS) through uplink data communication \cite{1,2}. However, nowadays the unlicensed industrial, scientific, and medical (ISM) bands are commonly used by most of the UAVs for point-to-point communication that limits their future applications due to the limited transmission range, rate, and connectivity.
\par
Recently, the researchers have shown significant interest in cellular-connected UAV communication that provides better performance than the point-to-point communication \cite{3}. Moreover, the feasibility tests of the cellular-connected UAVs carried out in the field trials are successful \cite{4,5}. Cellular-connected UAV communication is considered in two different kinds of scenarios. Firstly, a UAV could be deployed and used as an aerial BS or access point (AP) or relay to assist the ground deployed static users. Secondly, UAVs could be connected to a cellular network as aerial users and served through advanced cellular technologies such as Long Term Evolution (LTE) and 5G. Most of the recent literature \cite{6,7,8} has considered the former scenario. However, the research work with the latter scenario is rare, especially considering a large number of aerial users. 
\par
Currently, there are around half-a-million commercial drones and more than 1.5 million recreational drones registered with the federal aviation administration (FAA) in the United States. It is expected that the number of drones will double in 2024 \cite{9}. Additionally, it is observed that a large number of UAVs are required during disastrous situations to deliver aid or food supply to the affected people. Moreover, a large number of UAVs deployment for the intelligence, surveillance and security is very common these days, especially during war. Furthermore, most of the emerging applications are benefiting from the cellular-connected UAV networks to enhance communication. These applications include live video streaming and virtual reality (VR) etc., that require high communication rates for uplink data transmission from UAVs to ground BS known as GBS \cite{10}. However, due to the limited battery power of the UAVs, the requirement to fulfill their high transmit data rates in an energy efficient way is a challenging task in a cellular-connected UAV network \cite{11}. Furthermore, it becomes more challenging to provide connectivity to the large number of UAVs with diverse quality of service (QoS) requirements \cite{12}.
\par
\begin{figure*}[!ht]
	\centering
	\includegraphics[scale=0.6]{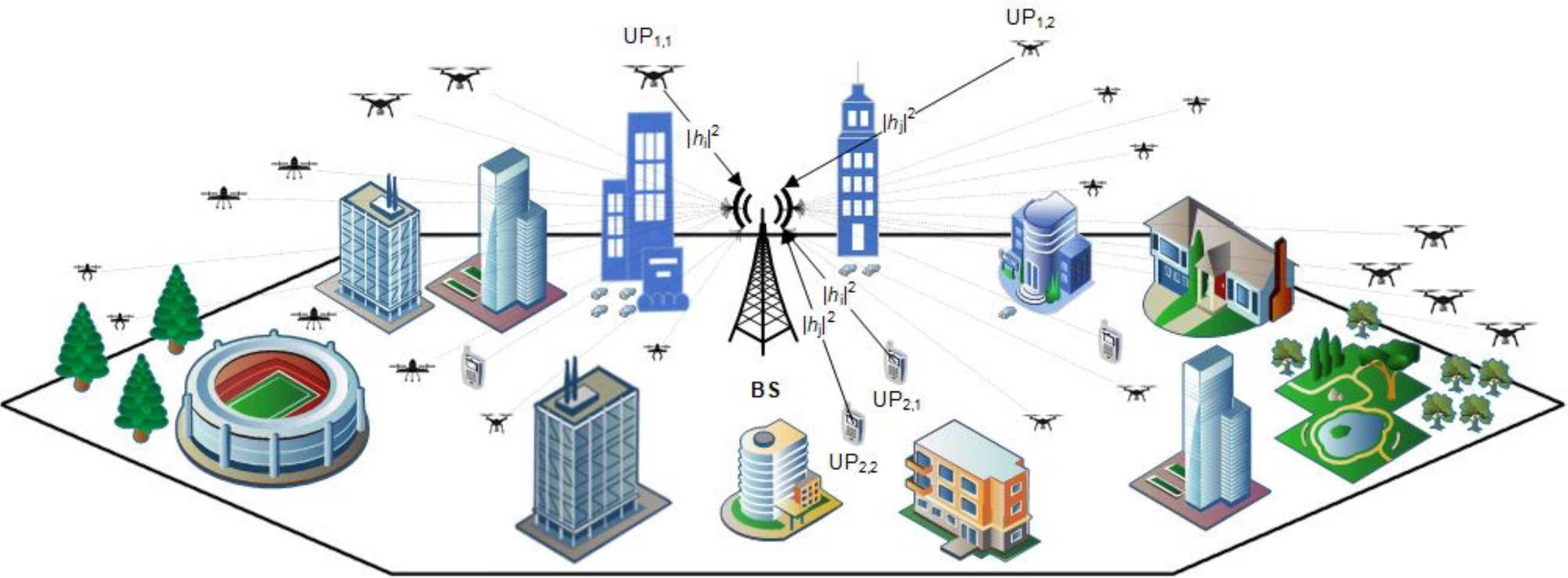}
	\caption{Cellular-connected mobile aerial users in a NOMA UAV network.}
	\label{Figure:1}
\end{figure*}
Non-orthogonal multiple access (NOMA) is considered as a potential candidate and a key enabler technique for B5G due to its support for the massive connectivity \cite{13,14,15}. There are multiple sub-classes of NOMA \cite{15a,15b,15c}, but power domain-NOMA (PD-NOMA) is well exploited in literature. In PD-NOMA, multiple users are paired together and served over the same radio resource block (RB), where BS allocates different power levels to these paired users to facilitate efficient data recovery through successive interference cancellation (SIC) \cite{15d}. According to uplink PD-NOMA,the BS first decodes the high power signals of strong channel users by treating low power signals from weak users as noise. This is followed by SIC to recover the low power signals of weak users from the superimposed signal. Prior research has thoroughly investigated the user pairing and power allocation techniques for both downlink and uplink NOMA \cite{16,17,18,19}. Moreover, recently the issues of NOMA user pairing under mobility constraint are investigated in \cite{20}. That leads us to further investigate the mobility in UAV assisted NOMA networks. In some of the recent research works on UAV assisted NOMA \cite{21}, the energy efficient placement of a mobile UAV-BS is performed to guarantee the QoS of the static ground users. Additionally, the authors investigated the effect of diverse user pairing schemes on the energy efficiency of the mobile UAV-BS. Furthermore, an energy efficient trajectory optimization technique is proposed in \cite{22} for a mobile UAV considering the point-to-point wireless communication. Similarly, in \cite{23} and \cite{24}, the trajectory of a mobile UAV-BS was optimized to maximize the users rates for downlink and uplink NOMA respectively. Additionally, there are some papers like \cite{25,26}, where UAVs are used for air-to-ground (A2G) cooperative NOMA (CNOMA) to assist uplink transmissions. 
\subsection{Motivation and Contributions}
Most of the previous studies show that the UAV is mostly used as a BS or a relay to assist cellular networks and little research has been conducted on the cellular-connected UAVs as mobile aerial users. Moreover, the investigations on energy efficient UAV pairing and power optimization of largely deployed mobile UAVs as aerial users are still an open issue. Considering this, and the research challenges discussed earlier, has motivated us to investigate cellular-connected UAVs as mobile aerial users. Hence, unlike existing literature, this work exploits NOMA in the cellular-connected UAV network, where UAVs are considered as mobile NOMA users. In particular, there are three primary objectives of this work: (i) to investigate and understand the channel variations of the mobile UAVs required for pairing and power allocation, (ii) to minimize the energy consumption of mobile UAVs for uplink NOMA transmission through power optimization, and (iii) to guarantee the QoS requirements of the mobile UAVs through optimal pairing. The overall contributions of this paper are summarized as follows:
\begin{itemize}
	\item Firstly, the main goal is formulated as a joint optimization problem for pairing and power allocation of mobile UAVs under uplink NOMA protocol. 
	\item The main problem is then divided into two sub-problems, where a power optimization algorithm is initially used to find optimal transmit powers that guarantee the required individual target rates of the mobile UAVs for uplink NOMA transmission. Then, a two-sided matching algorithm is developed to efficiently solve the combinatorial user pairing problem and to perform energy efficient pairing under minimum and maximum power constraints of the mobile UAVs. The goal is to guarantee QoS provision to the mobile UAVs by optimal pairing and to minimize the energy consumption of the mobile UAVs by allocating optimal transmit powers.
	\item Finally, the pairing complexity and performance of the proposed technique is analyzed. Numerical and simulation analysis is performed to show the achieved gains of the proposed technique compared to the conventional NOMA user pairing and power allocation techniques in terms of target rate and energy efficiency.
\end{itemize}
\par
The rest of the paper is organized as follows. Section II presents the system model. The proposed technique is presented in Section III. The pairing complexity analysis is discussed in section IV. In section V, simulation and numerical results are presented to show the effectiveness of our proposed technique. Section VI presents some potential future directions. Section VII concludes the paper.

\section{System Model}
Consider a UAV network with $K_T$ cellular-connected mobile UAVs as aerial users, where $\mathbb{K_T}$ = $\{1,2,...,K_T\}$. All UAVs belongs to set $\mathbb{K_T}$ in the network are transmitting their data to a BS through uplink NOMA transmission protocol as shown in Fig.\ref{Figure:1}. The cellular-connected mobile UAVs are moving within a cell boundary with limited battery power constraints. Two UAVs\footnote{In this paper, it is considered that two UAVs will be paired over any particular $m^{th}$ subchannel to reduce the pairing complexity as well as to control the error propagation of SIC, which becomes more prominent when more than two users are paired in conventional NOMA.} as a NOMA pair can transmit their data using a shared channel resource and distinct power levels according to the uplink NOMA principle. Accordingly, the bandwidth $B$ is equally divided into $M=\frac{K_T}{2}$ subchannels. Moreover, some further details about the set $\mathbb{K_T}$ are discussed in \textit{Remark-1}.
\par
\paragraph*{Remark-1}
Our preliminary goal is to analyze the performance (in terms of energy efficiency and required data rate) of mobile UAVs using NOMA protocol. Therefore, while a cellular system may have both ground and aerial users simultaneously served through a BS, we specifically consider a case where the network contains only mobile UAVs as cellular-connected aerial users i.e., $\mathbb{K_T}$=$\mathbb{K_A}$, where $\mathbb{K_A}$ represents UAVs. In other words, as shown in Fig.\ref{Figure:1}, it can be considered that the ground users $\mathbb{K_G}$ are already paired ($UP_{2,1}, UP_{2,2}$) on separate subchannels that can not be shared with the cellular-connected aerial users ($UP_{1,1}, UP_{1,2}$). While the proposed technique can be extended to mixed pairing between mobile aerial and ground users, such an approach is subject to our future research works.

\subsection{Channel and Mobility Model}
To demonstrate the mobility scenario, a random waypoint mobility (RWP) model \cite{27} with 3D positions $(x,y,H)$ of the mobile UAVs at a constant altitude $H$ is considered, while the ground BS is fixed at the center $(x_0,y_0)$ of a cell. Further, the UAVs due to mobility change their angle of direction, speed, and destination point within the specified cell boundary at each time instant ($t$). Moreover, it is considered that the Doppler effect due to UAV mobility is perfectly compensated at the BS \cite{27a}. Due to the distance variation between the BS and mobile UAVs, channel ranking lists of mobile UAVs at each new position are updated at the BS for each time instant $t$. Considering, a constant height $H$, the time variant distance between the BS and a $UAV_k$ can be given as

\begin{equation}
\label{eq01}
d_k(t) = \sqrt{D_k^2(t) + H^2},\quad \forall k \in \mathbb{K_T},
\end{equation}

where $D_k(t)$ is the horizontal distance $\sqrt{(x_k-x_0)^2 + (x_k-x_0)^2}$ between a $UAV_k$ and BS at time instant $t$.
\par
Generally, the A2G links are considered as line of sight (LOS)-dominated channels. According to the 3rd Generation Partnership Project (3GPP),  a certain height threshold is required to achieve LOS probability \cite{28}. As mobile UAVs are considered in a dense urban region, the probability of non-LOS (NLOS) can not be easily ignored \cite{29}. Therefore, the elevation angle dependent probabilistic LOS channel model \cite{2},\cite{30} is used where channel probability of a mobile UAV is determined by both LOS and NLOS channel probabilities are expressed as 

\begin{equation}
\label{eq02}
Pr_{k}^{LOS}(t)=\frac{1}{1+ \zeta \exp (-\delta [ \theta_k -\zeta] )) }, 
\end{equation}

\begin{equation}
	\label{eq03}
Pr_{k}^{NLOS}(t)=1-Pr_{k,LOS}(t),
\end{equation}

where $\zeta$ and $\delta$ are the environment (dense urban, urban, rural) constants, while $\theta_k=\frac{180}{\pi} \times \arcsin \left(  \frac{H}{d_k} \right)$ represents the angle of elevation between the mobile UAV and BS. 

Furthermore, there are some loss factors ($\Lambda_{LOS}$,$\Lambda_{NLOS}$) involved during the signal transmission via A2G links. The path loss incurred during signal transmission via LOS and NLOS A2G links is given as 

\begin{equation}
	\label{eq033}
	PL_{k}^{LOS}(t)=10\psi\log(d_k)+\Lambda_{LOS},
\end{equation}

\begin{equation}
	\label{eq0333}
	PL_{k}^{NLOS}(t)=10\psi\log(d_k)+\Lambda_{NLOS},
\end{equation}

where $\psi$ represents the path loss exponent \cite{21,31}. Thus, the combined path loss for A2G link of a $k^{th}$ mobile UAV, considering both LOS and NLOS links, can be expressed as \cite{30,31}


\begin{equation}
	\small
	\label{eq04}
	PL_k(t)=Pr_{k}^{LOS}(t) PL_{k}^{LOS}(t) +Pr_{k}^{NLOS}(t)PL_{k}^{NLOS}(t).
\end{equation}

\begin{table}[]
	\centering
	\caption{List of symbols used in this paper.}\label{tb22}
	\begin{tabular}{|l|l|}
		\hline
		\textbf{Symbols} & \textbf{Definition} \\
		\hline
		$K_T$ &  Number of UAVs\\
		\hline
		$H$ & Height of a UAV from the ground \\
		\hline
		$\theta$ & Angle of elevation \\
		\hline
		$Pr_{i}^{LOS}$ & Probability of a LOS link \\
		\hline
		$Pr_{i}^{NLOS}$ & Probability of a NLOS link \\
		\hline
		$PL$ & Combined path loss \\
		\hline
		$\psi$ & Path loss exponent \\
		\hline
		$T_T$ & Total flight duration of a mobile UAV \\
		\hline
		$d_i$ & The distance between the BS and a mobile $UAV_i$ \\
		\hline
		$h_i$ & Channel coefficient of a mobile $UAV_i$  \\
		\hline
		$R_i$ & Data rate of a mobile $UAV_i$\\
		\hline
		$R_T^i$ & Target rate of a mobile $UAV_i$\\
		\hline
		$CH_{th}$ & Channel threshold \\
		\hline
		$P_i$ & Transmit power of a mobile $UAV_i$ \\
		\hline
		$P_{r,i}$ & Received power of a mobile $UAV_i$ \\
		\hline
		$P_c$ & Communication power \\
		\hline
		$\alpha$ & Power allocation factor \\
		\hline
		$\eta_{EE}^i$ & Energy efficiency of a mobile $UAV_i$ \\
		\hline
		$E_T$ & Total energy consumption \\
		\hline
		$\kappa$ & Number of UAVs with successful transmission \\
		\hline
	\end{tabular}
\end{table}

\begin{figure*}[htp]
	
	\centering
	\includegraphics[width=.25\textwidth]{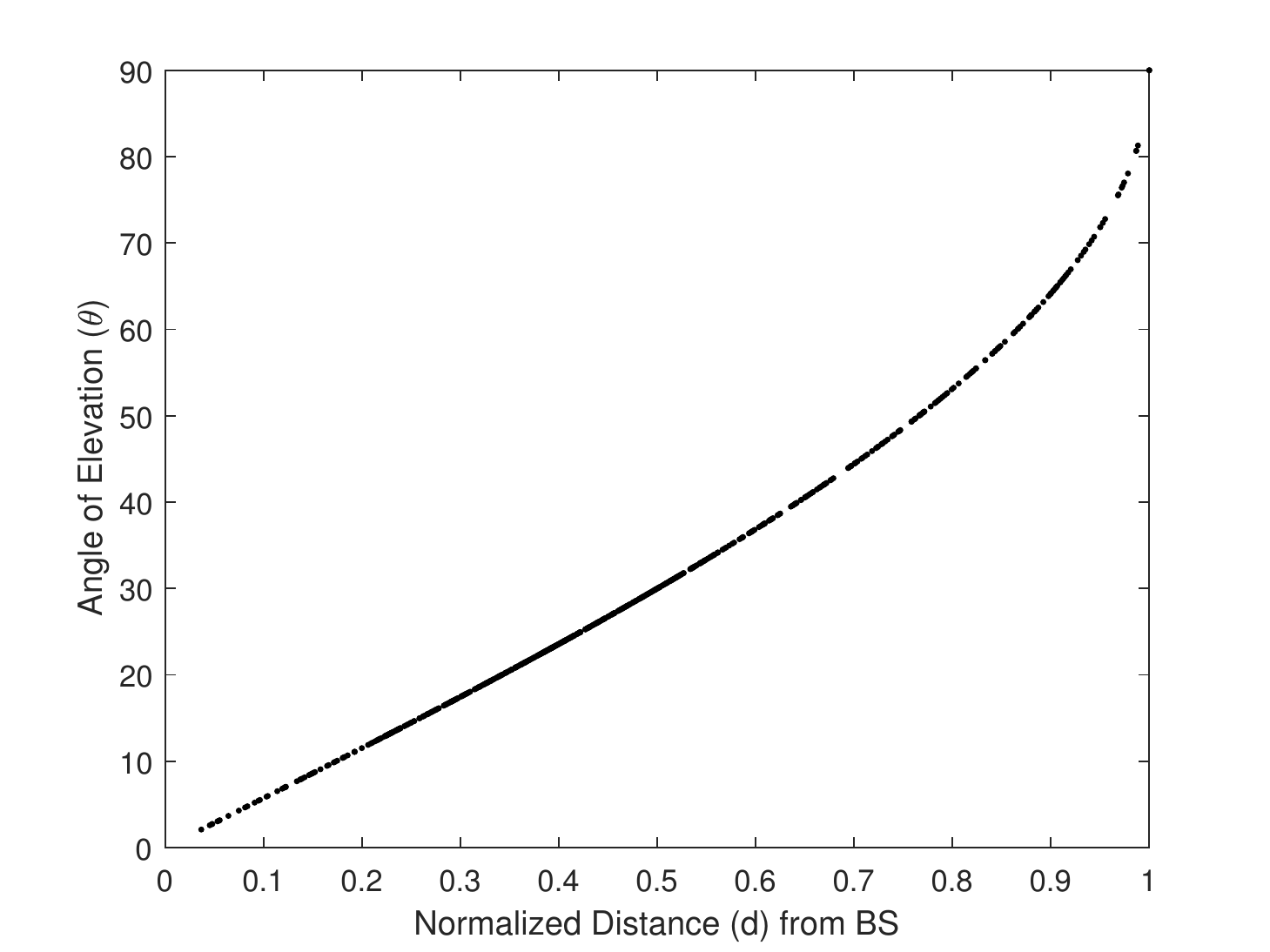}\hfill
	\includegraphics[width=.25\textwidth]{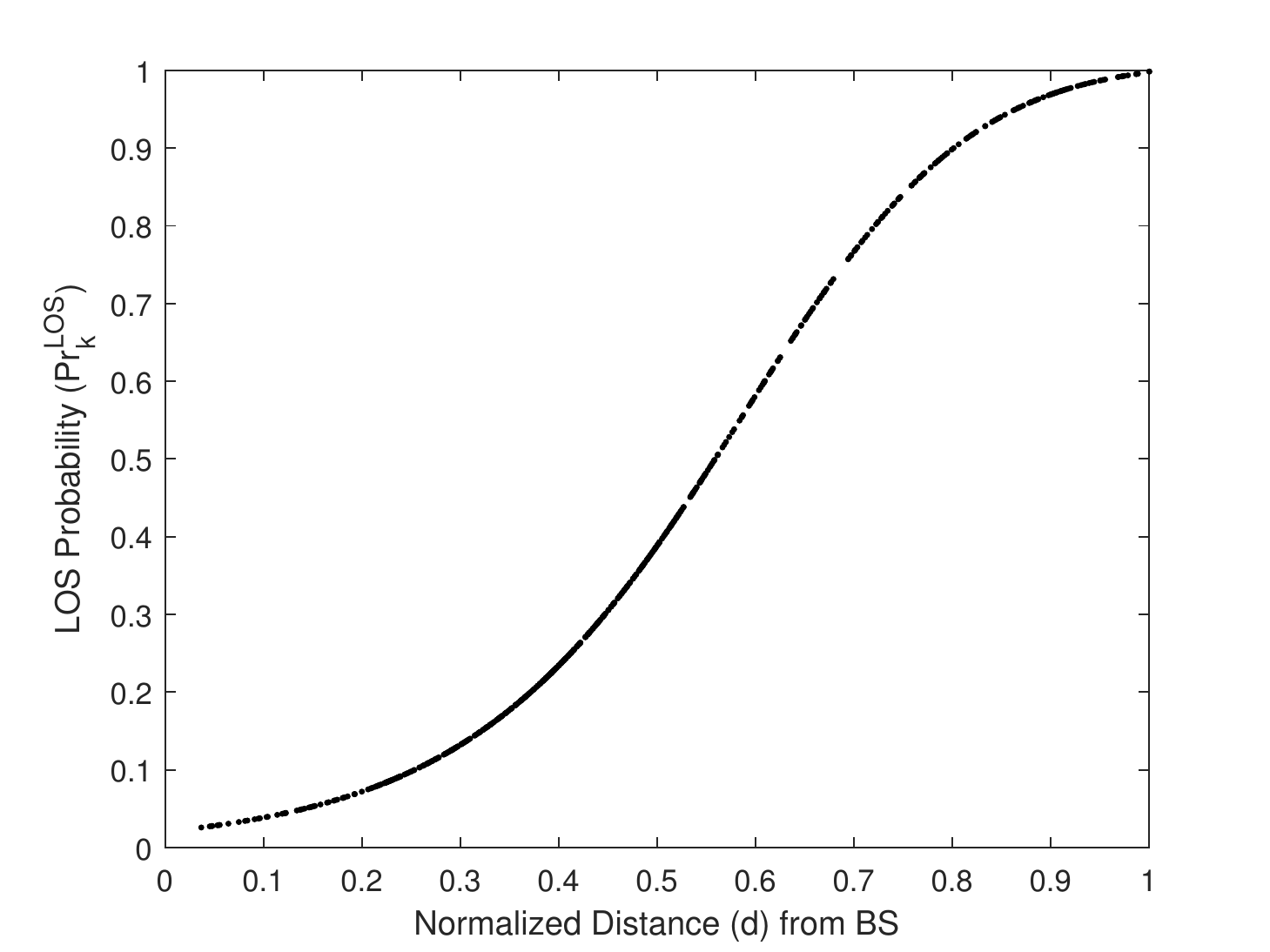}\hfill
	\includegraphics[width=.25\textwidth]{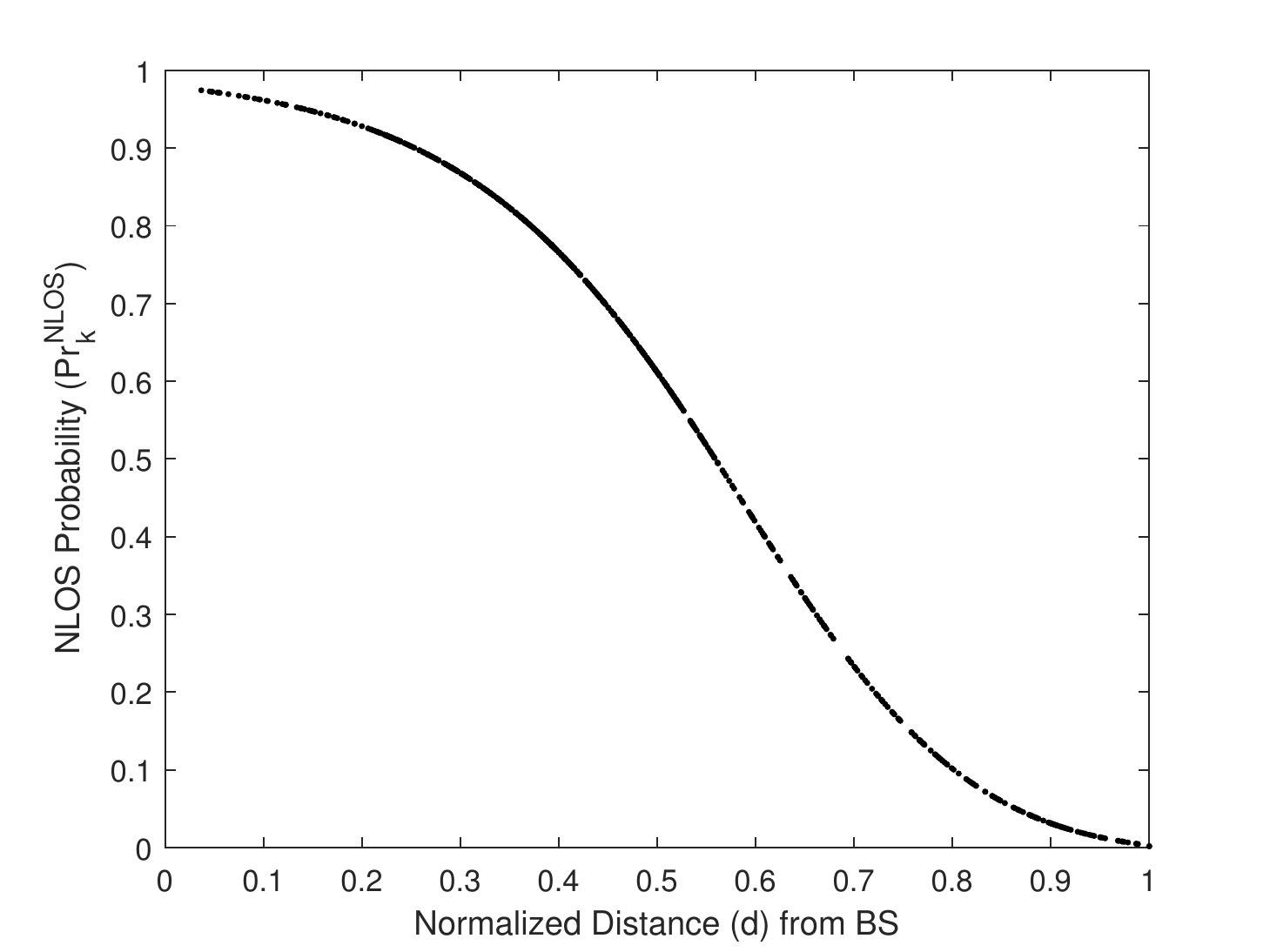}\hfill
	\includegraphics[width=.25\textwidth]{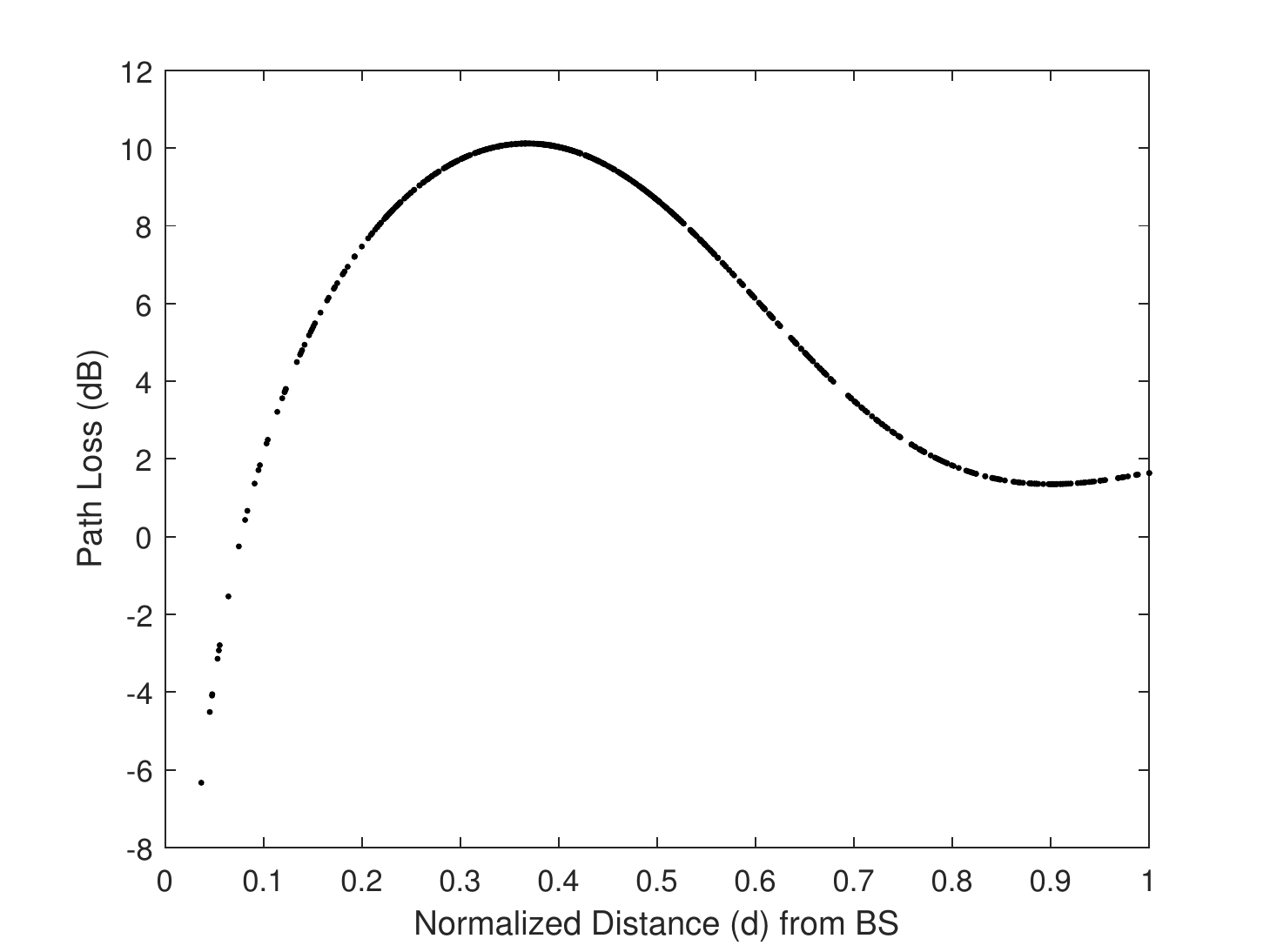}
	
	\caption{Collective channel variations of mobile UAVs ($K_T$=50) during their total flight duration ($T_T=120s$) is shown at normalized distances from the BS. The affect of angle of elevation, LOS and NLOS probabilities is shown on combined path loss using figures from left to right, respectively.}
	\label{fig:figure3}
	
\end{figure*}

Additionally, due to the continuously varying distances of the mobile UAVs from the BS, their channel gains also vary. Therefore, the channel coefficient ($h_k$) from a mobile UAV to BS is given by

\begin{equation}
	\label{eq05}
	h_k(t)= PL_k(t) \tilde{h}_k(t),~~k=\{1,...,K_T\},
\end{equation}

where $\tilde{h}_k(t)$ is a complex valued random variable with $\mathbb{E}[|\tilde{h}_k(t)|^2]$ = 1 corresponds to the small-scale Rayleigh fading coefficient, while $PL_k(t)$ corresponds to the large-scale fading that is related to the probabilities of LOS and NLOS links given in Eq. \ref{eq04}.

Without loss of generality, the channel gains follow $|h_1|^2 \geq |h_2|^2 \geq \ldots \geq |h_k|^2 \geq \ldots \geq |h_K|^2$ where lesser the subscript number means the higher channel gain. For uplink NOMA transmission, two paired mobile UAVs ($i,j$) with channel gain  $|h_i|^2 > |h_j|^2$ can share the same subchannel while the transmit powers of paired UAVs will be different using the PA factors of $\alpha_i$ and $\alpha_j$, where $\alpha_i > \alpha_j$. 
\par
\paragraph*{Assumption-1}
For ease of exposition, perfect CSI and channel rankings of the mobile UAVs are assumed to be known at the BS.

\subsection{Achievable Rates of UAVs under Mobility}
It is considered that the mobile UAVs have limited battery power and each $k^{th}$ mobile UAV can utilize its transmit power in the range between $\{P_{k,min},P_{k,max}\}$ to transmit a message $x_k$ to the BS. Therefore, considering $K=2$ UAVs in a pair, the received signal at the BS is given as

\begin{equation}
	\label{eq07}
	Y = \sum_{k=1}^{K} h_k \sqrt{P_{r,k}} x_k + n,
\end{equation}

where $h_k$ is the instantaneous channel gain of the $k^{th}$ mobile UAV, $n$ represents the additive white Gaussian noise (AWGN) at the BS, and $P_{r,k}$ is the received power at time instant ($t$) of the $k^{th}$ mobile UAV, given as \cite{21},\cite{32}

\begin{equation}
	\label{eq077}
	P_{r,k}(t) = \alpha_k P_k - PL_k(t),
\end{equation}

where $\alpha_k$ is the power allocation (PA) factor and $P_k$ is the total transmit power of a $UAV_k~(k~\in~K_T$). 

\paragraph*{Assumption-2}
In the analysis of the achievable rates for uplink NOMA transmission, it is assumed that the SIC process at the BS is perfect.  
\par
According to the \textit{Assumption-2}, and considering $K=2$ UAVs in a pair, where the variables $UAV_i$ and $UAV_j$ denote the strong and weak channel users, the instantaneously received signal-to-interference-plus-noise-ratio (SINR) of the weak channel mobile $UAV_j$ at the BS while utilizing Eq. (\ref{eq077}) is given as

\begin{equation}
	\label{eq08}
	\rho_j(t) =\frac{\alpha_j P_{j} |h_j|^2 }{N_o} .
\end{equation}

Corresponding achievable rate of a weak channel mobile $UAV_j$ during its total flight duration $T_T$ is given as a function of a mobile UAV flight time $T_T(t)$, presented as

\begin{equation}
	\label{eq09}
	R_j(T_T(t)) = \int_{0}^{T_T} \log_2 \left( 1+\frac{\alpha_j P_{j} |h_j|^2 }{N_o} \right) dt,
\end{equation}

where, $N_o$ represents variance of AWGN. Similarly, for a better channel $UAV_i$, the received SINR at the BS at time instant $t$ is given by

\begin{equation}
	\label{eq10}
	\rho_i(t) =  \frac{\alpha_i P_{i}(t) |h_i|^2 }{\alpha_j P_{j}(t) |h_j|^2 + 1},
\end{equation}

where the term in the denominator represents interference from the weak channel $UAV_j$. Similarly, the achievable rate of strong channel mobile $UAV_i$ is given as

\begin{equation}
	\label{eq11}
	R_i(T_T(t)) = \int_{0}^{T_T} \log_2 \left( 1+ \rho_i(t) \right) dt.
\end{equation}
\par
Further, it is considered that the target rate $R_k^T$ of a mobile $UAV_k$ is its orthogonal multiple access (OMA) rate, given as \cite{16}:

\begin{equation}
	\label{eq12}
	R_k^T(T_T(t)) = \int_{0}^{T_T} \frac{1}{2} \log_2 \left( 1+ \Theta_k |h_k|^2  \right) dt,
\end{equation}

where $\Theta_k=P_{k}/N_o$.
\par
\paragraph*{Remark-2}
Here the OMA rates are considered as the target rates of the mobile UAVs. It is to be noted that OMA results in mutiplexing loss of $1/2$, specially if the UAVs are served in the time division multiple access (TDMA) manner. 

\subsection{Energy Model for Mobile UAVs}
There are many factors involved in the energy consumption of a UAV such as communication energy for data transmission and reception due to signal radiation and processing, flying energy to keep UAV mobile, and sometimes hovering around if required \cite{10},\cite{22}. The flying energy of UAVs depends on the speed and acceleration while the communication energy of the UAVs depends on the data transmission/reception. Further, the amount of communication energy is substantial and have an impact on the flying time and network lifetime. Additionally, the energy consumption of a UAV continuously transmitting data to the BS in real time or mission critical systems is very high. Therefore, energy efficient data transmission is required where the UAVs can transmit with the minimum power to fulfill the required data rates.

Let the total energy of a UAV be denoted by $E_T$, which is the combination of communication energy and flying energy $e_c$, $e_f$ respectively. Numerically, it can be written as \cite{10,22}
\begin{equation}
\label{eq06}
E_T(T_T(t)) \approx \int_{0}^{T_T} \left( e_c(t) + e_f(t) \right) dt,
\end{equation}

\begin{equation}
	\label{eq066}
	E_T(T_T(t)) = \int_{0}^{T_T} \left( P_{c}(utx(t)) + e_f(t) \right) dt,
\end{equation}

where $e_c(t)= \int_{0}^{T_T} P_{c}(utx(t)) dt$, factor $T_T$ represents total flight duration and $P_c$ is the power required for a single uplink data transmission $utx(t)$ at time ($t$).
\par
\paragraph*{Assumption-3}  In order to analyze the impact of communication energy of a mobile UAV, it is assumed that $e_f$ is constant term whose closed-form expression and dependent variables are given in most of the previous works \cite{10,22,33}. However, $P_c$ of a mobile UAV is considered variable and depends on the proposed power allocation scheme used to achieve the required target rate ($R_T$).

\section{Pairing and Power Optimization for Cellular-Connected Mobile UAVs}


\subsection{Problem Definition and Working of the Proposed Technique}
In general, previous literature on NOMA \cite{16,17,18,19} consider fixed pairing of the static users that does not work well when applied in a real time, particularly pairing of mobile users with continuous channel gain variations. An example from the above problem is NOMA principle violation problem (NPVP) for two mobile users presented in \cite{20}, where NPVP represents scenarios where the paired near and far users come very close to each other or their channel ordering changes due to mobility. In \cite{20}, rather than breaking the pairs and making new pairs, the pairing performed at the start is maintained even if the NPVP problem arises by adjusting the power allocations or switching the roles of near and far users to reduce the complexity of repetitive pairing/unpairing of mobile users.

Similarly, several issues arise while considering the fixed pairing techniques especially in the NOMA UAV networks where the users are mobile aerial vehicles. These pairing schemes mostly rely on making a channel ranking based list of candidate users, and then make pairs between the strong and weak channel users. If we employ them in the considered mobile UAV network of $K_T$ users, in each iteration, even if the channel gains of a couple of users vary, the whole ranking list changes, and correspondingly all user pairings will be continuously affected. For these existing pairing schemes, the following issues arise:
\begin{itemize}
    \item Pairing and unpairing of all $K_T$ mobile UAVs at each time instant ($t$) during mobility is more complex and cumbersome at the BS; its complexity becomes too high.
    \item Furthermore, breaking and pairing of all mobile UAV pairs at each time ($t$) will result in frequent handover failures of mobile UAVs and lead to a poor QoS.
    \item Additionally, the fixed pairing technique unpairs all of the $K_T$ mobile UAVs at each time instant ($t$) even if the target rates of the mobile UAVs are satisfied due to having a direct LOS links with some small channel variations.
    \item Also, the users in middle of the ranking list i.e., close to the boundary of the strong and weak user regions in the list are vulnerable to more problems. A boundary user, considered in the strong channel users region in the ranking list in one time slot, may fall into the weak users region in the next slot, and may be required to transmit with totally different power according to uplink NOMA protocol; such transmit power variations may cause issues at the users' ends.
    \item Moreover, the fixed pairing or power allocation techniques are mostly rate, outage or energy oriented, but do not care about the channel gain or power difference threshold that should be maintained for a successful data recovery of users' messages at the BS. Therefore, the data of most of the mobile UAVs cannot be successfully recovered at the BS in uplink NOMA or at the users' ends in downlink NOMA \cite{18,38}.
\end{itemize}
\par
There are many possible solutions for the above discussed problems and the potential solution depends on the mobility profile of the UAVs in the network, their rate constraints, and other target goals. For example, a heterogeneous UAV network contains UAVs with all three types of mobility profiles namely low mobility, average mobility and high mobility. While a homogeneous UAV network contains only a single type of mobile UAVs among those three types, developing a pairing solution for a heterogeneous UAV network is more challenging than a homogeneous network. Moreover, thanks to the LOS dominated link of the mobile UAV that is not frequently changing during the flight of the mobile UAVs at certain height ($H$) \cite{28} and reduces the pairing complexity at the BS (for details see \textit{Remark-4}).
\par
\paragraph*{Remark-4}
In NOMA, a BS makes user pairs according to the channel gain differences between them. If it is less than the defined threshold, the BS breaks that pair and makes a new pair. This breaking and pairing of users can increase the complexity at the BS especially if the channel gains are continuously changing due to mobility \cite{20}. The details of pairing complexity for mobile UAVs scenario is discussed in section IV.
\par
\begin{figure}[t!]
	\centering
	\includegraphics[scale=0.4]{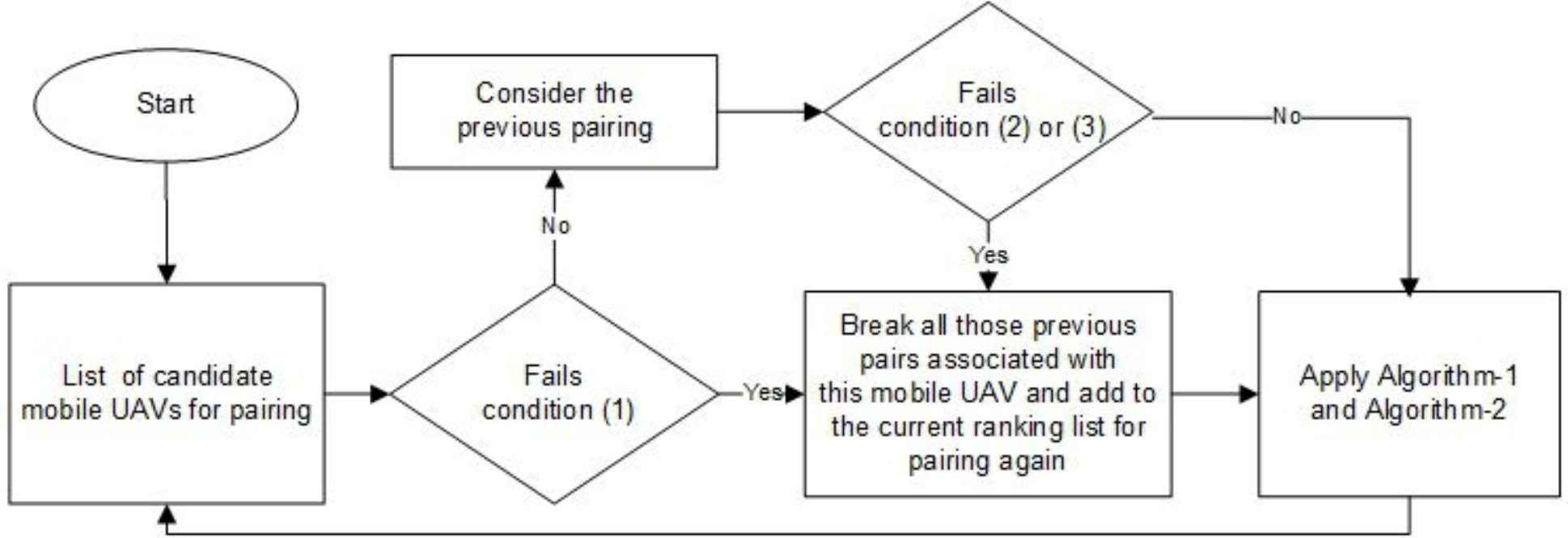}
	\caption{Working principle of the proposed technique.}
	\label{Figure:111}
\end{figure}

Therefore, in this paper, we propose an energy efficient and low complexity pairing solution for a homogeneous UAV network  where pairing of the mobile UAVs is performed in real time by observing the mobile UAVs. The working flow of the proposed pairing technique is shown in Fig.\ref{Figure:111} and it maintains the pairing of mobile UAVs until any one of the following conditions is violated:
\begin{enumerate}
    \item the channel gain of the mobile UAVs becomes too high or too low, 
    \item the minimum target rate constraint of the mobile UAVs is violated or 
    \item the minimum power difference constraint between two mobile UAVs in a pair is violated.
\end{enumerate}
Particularly, the pairing will be performed only for the mobile UAVs that violate one of these constraints. The ultimate goal is to satisfy the QoS rate requirement of each mobile UAV with minimum power consumption by pairing mobile UAVs based on their channel ranking list at the BS. However, for the heterogeneous UAV networks the potential problems and recommendations are presented in section VI and they are considered as our future research work.
\par
\subsection{Problem Formulation}
Consider a homogeneous UAV network where $K_T$ mobile UAVs with instantaneous channel gains, asymmetric QoS requirements and limited power constraints need to be paired using uplink NOMA principle. Two mobile UAVs per pair are considered (according to \textit{Remark-1}) for a successful uplink NOMA transmission, and with a goal that their required individual target rates should be achieved with minimum power consumption. The overall optimization problem can be expressed as


\begin{subequations} \label{problem}
\begin{alignat}{2}
& \max\limits_{T_T,P,H}
& & ~\eta_{EE} \label{eq:15a}\\ 
& ~ \text{s.~t.~}& & R_k \geq R_k^T, \quad k=\{1,...,K_T\},~~\forall k \in \mathbb{K_T}, \label{eq:15b}\\
& & &  P_{k,min} \leq  P_k \leq P_{k,max},\quad k=\{1,...,K_T\} \label{eq:15c}\\
& & &  P_{i} > P_{j},~\text{if}~|h_{i}|^2 > |h_{j}|^2,   \label{eq:15d}\\
& & &  P_i-P_j \geq P_{th}, \label{eq:15f}\\
& & &  |h_i|^2-|h_{j}|^2 \geq Ch_{th},  \label{eq:15e}\\
& & & M =\sum_{i=1}^{I} \sum_{j=1}^{J} \omega,~~\forall i \in \mathbb{I},\forall j \in \mathbb{J}. \label{eq:15g}
\end{alignat}
\end{subequations}
where function (\ref{eq:15a}) maximizes the energy efficiency of mobile UAVs considering their target rate constraints in (\ref{eq:15b}) and power constraints for uplink NOMA transmission are given in   (\ref{eq:15c}),(\ref{eq:15d}) and (\ref{eq:15f}), where constraints in (\ref{eq:15e}) describes the channel conditions necessary for user pairing in uplink NOMA. Finally, the constraint is given in (\ref{eq:15g}) that sum of all matched pairs should be equal to the number of available subchannels ($M$). It will be discussed in detail in subsection-D.
\par
\subsection{Power Optimization for Mobile UAVs}
In this section, a power optimization algorithm is presented to find the minimum optimal powers of possible candidate UAVs for pairing before applying the matching based pairing algorithm. Unlike the previous matching based user pairing and power allocation technique \cite{34}, in the proposed technique minimum optimal powers of mobile UAVs are calculated  separately before user pairing. In the technique proposed in \cite{34}, the negotiation process between proposers and selectors for a suitable power allocation creates an extra overhead, especially when applied to mobile UAVs as in our case. Instead, a divide and conquer approach is used here, where the BS at first finds the minimum optimal powers of the possible candidate mobile UAVs for pairing by considering their required target rates and channel threshold constraints according to \textit{Remark-3}.
\par
\paragraph*{Remark-3} Note that, unlike the previous literature on matching based user pairing and power allocation for downlink NOMA \cite{34}, a different matching approach is presented due to the following reasons. Firstly, the previous matching based pairing technique used a combined distributive approach for the user pairing and power allocation to minimize the complexity at the BS which is not applicable to our considered uplink transmission model. Secondly, if the UAVs are mobile, then a dynammic matching technique \cite{35} is required, where preferences of the mobile UAVs change after a certain time interval, and it is more suitable to be performed at the BS having more powerful processing and unlimited energy resources than the UAVs with limited resources. Thirdly, the proposed power allocation and matching based pairing technique considers only possible candidate UAVs for pairing at each time instance unlike the exhaustive search technique that considers all users for pairing and is impractical due to high complexity. Lastly, the proposed technique is developed as a centralized approach because if the BS has all the channel ranking information of the mobile UAVs then performing power optimization and matching based pairing is more suitable for uplink NOMA transmissions.

\begin{algorithm}[t!]
	\small
	\caption{Solution to Problem (P1)}
	
	\begin{algorithmic}[1]
		\State \textbf{Generate} UAVs Coordinates $\triangleright~ \footnotesize \text{using RWP mobility model}$
		\State ~~~~~~~~~~~~\{\textbf{return} $d$, $T_T$\}
		\State \textbf{Compute} Channel Matrix $\textbf{H}$ $\triangleright~ \footnotesize \text{Known at BS}$
		\State ~~~~~~~~~~~~\{\textbf{return} $rankinglist\{1,\dots,\bar{K}_T\}$ \} $\triangleright~ \footnotesize \text{ascending sorted}$
		\For {each time instant $t$}
		\For {each UAV from $rankinglist_j$}
		\State $UAV(j) \leftarrow rankinglist_j$
		\For {$i \leftarrow 1, rankinglist_i$}
		\State $i \neq j$
		\If {$|h_i|^2-|h_j|^2 \leq Ch_{th}$}
		\Repeat
		\State $Set~P_{j,curr}\leftarrow (P_{j,min}+P_{j,max})/2$
		\State \textbf{Calculate} Current Rate $R_j$ $\triangleright~ \footnotesize \text{using Eq.(\ref{eq09}),(\ref{eq11})}$
		\If {$R_j < R_j^T$}
		\State $P_{j,min} \leftarrow P_{j,curr}$
		\Else
		\State $P_{j,max} \leftarrow P_{j,curr}$
		\EndIf
		\Until Eq.(\ref{eq:16b}), Eq.(\ref{eq:16c})
		\State $P(j,i)=P_{j,curr}$
		\EndIf
		\EndFor
		\EndFor
		\EndFor
	\end{algorithmic}
\end{algorithm}

Generally, exhaustive search power allocation (ESPA) methods are used to find the minimum optimal transmit powers that guarantee the required target rates of NOMA users. However, the computational cost of such power allocation methods are very high. In this paper, a fast convergence power optimization algorithm is proposed to find minimum optimal transmit powers of the mobile UAVs that guarantee the required target rate without performing the exhaustive search. Instead, the working of the proposed power optimization algorithm is based on the binary search power optimization algorithm \cite{35a} that is much faster than ESPA.
\par
Furthermore, the above problem is joint power optimization and pairing problem which is difficult to be directly solved \cite{21}. Therefore, the power optimization part of the main problem as (\textbf{P1}) is given below

\begin{subequations} \label{problem2}
\begin{alignat}{2}
(\textbf{P1}):& \min\limits_{P,T_T}
& & (E_T(T_T)) \label{eq:16a}\\ 
& ~ \text{s.~t.~}& & R_k \geq R_k^T \quad \forall~k=\{1,2,...,K_T\} \label{eq:16b} \\
& & &  \alpha_i > \alpha_{j}, ~ \text{if}~|h_i|^2 > |h_{j}|^2 \label{eq:16c}\\
& & &  |h_i|^2-|h_{j}|^2 \geq Ch_{th},  \label{eq:16d}
\end{alignat}
\end{subequations}

The optimization function in (\ref{eq:16a}) minimizes the energy consumption of mobile UAVs during their flight time $T_T$, where (\ref{eq:16b}) is the target rate condition i.e., individual rate of a UAV should be greater than the target rate $R_T$, (\ref{eq:16c}) represents the power rule for uplink NOMA which states that power allocation factor for the strong channel UAV should be greater than the weak channel UAV and (\ref{eq:16d}) represents the channel difference constraint required for pairing in NOMA.
\par
Considering, the above constraints (\ref{eq:16b}) and (\ref{eq:16c}), defined in the problem (\textbf{P1}), the proposed solution is presented as (\textbf{Algorithm.1}), where minimum optimal powers of all mobile UAVs are searched to minimize their total energy consumption throughout their flight duration $T_T$.
\par
Initially, a ranking list ($rankinglist$) of all mobile UAVs is generated based on the available CSI at the BS so that the mobile UAV with weakest channel condition is at the top of the list followed by a mobile UAV with the better channel and so on. Please note that the list keeps updating in the later time instances i.e., $rankinglist \{1,\dots,\bar{K}_T\}$, and only contains the UAVs whose previous pairing needs changing or any new UAV candidates for pairing. Then the $rankinglist$ is divided into sets of  weak and strong channel gain UAVs, given as $rankinglist_i\{1,\dots,\frac{\bar{K}_T}{2}\}$ and $rankinglist_j\{\frac{\bar{K}_T}{2}+1,\dots,\bar{K}_T\}$, respectively. Here, the $rankinglist$ is in ascending order to find the minimum optimal powers of the weak channel mobile UAVs at first because according to the \textit{Assumption-3}, the BS performs perfect SIC for the UAVs with weak channels. Thus, it is quick and easy to resolve the constraints for the weak channel mobile UAVs at first. Moreover, it is necessary because these optimal powers of the weak mobile UAVs are used to search the optimal powers of the strong mobile UAVs (see Eq. (\ref{eq11})). Furthermore, the power allocations of the strong and weak channel mobile UAVs are satisfied based on their individual target rates. For the strong mobile UAVs the powers of weak mobile UAVs are considered as interference (see Eq. (\ref{eq11})). Therefore, it requires more power for a strong mobile UAV to satisfy its target rate and a conditional loop is used in \textbf{Algorithm-1} to keep increasing the power of the strong mobile UAV until conditions (\ref{eq:16b}) and (\ref{eq:16b}) are satisfied. Finally, the power allocations of all mobile UAVs are stored in a power allocation matrix (\textbf{P}). Ultimately it reaches to the final goal that is performing the power optimization by keeping the required target rates and pairing the mobile UAVs based on their channel rankings. Therefore, a power optimization algorithm is applied that minimizes the energy consumption of the mobile UAVs, guarantees the required target rates, satisfies the channel difference condition and converges faster than the ESPA.

\subsection{Energy Efficient Pairing for Mobile UAVs}
After power optimization of the mobile UAVs, an energy efficient pairing is required with minimum pairing complexity at the BS. Before presenting the matching based problem for an energy efficient pairing, it is necessary to define the following terms:\\
\par
\paragraph*{Definition-1} A UAV communication is said to be energy efficient, if the required target rate ($R_T$) of a mobile UAV is successfully achieved using minimum transmit power ($P_c$) during its flight time ($T_T$).
\par
According to \cite{10,21,22}, energy efficiency can be numerically represented as the ratio of sum rate and the total energy consumption of the UAVs in a network. Therefore, using Eq. (\ref{eq09}), Eq. (\ref{eq11}),  and Eq. (\ref{eq066}), the energy efficiency of the mobile UAVs at time instant ($t$) can be expressed as

\begin{equation}
	\label{eq:18}
	\eta_{EE}(t) =  \sum_{k=1}^{K}   \frac{R_k(t)}{E_{T,k}(t)},  \quad ~k=\{1,\dots,K_T\}.
\end{equation}
\par
\paragraph*{Definition-2} A binary variable $\kappa(t)$ is defined to represent number of mobile UAVs whose current rate is greater than the required target rate and it is considered as successful uplink data transmission ($utx$) of mobile UAVs at time ($t$).

\begin{equation}
	\label{eq13}
	\kappa_k (t)=\begin{cases}
		1 \quad \quad &\text{if}~\, R_k(t) \geq R_k^T, \quad \forall k \in \mathbb{K_T},  \\
		0 \quad \quad &\text{if}~\, R_k(t) < R_k^T, \quad \forall k \in \mathbb{K_T},
	\end{cases}
\end{equation}

\begin{equation}
	\label{eq14}
	\sum_{k=1}^{K_T} \kappa_k(t) \leq K_T , \quad k=\{1,\dots,K_T\},
\end{equation}

where Eq. (\ref{eq14}) represents total number of mobile UAVs whose QoS rate constraint (according to \textit{Remark-2}) is successfully achieved at time instant ($t$).\\
\par
\paragraph*{Definition-3} An ($I \times J$) matrix $\mathbb{W}$ is defined to store the matched pairs of strong mobile UAVs ($I$) with weak mobile UAVs ($J$) at time instant ($t$) using the following corresponding entries

\begin{equation}
	\label{eq133}
	\omega(i,j)(t)=\begin{cases}
		1 \quad \quad &\text{if}~\, matched, \\
		0 \quad \quad &~\, otherwise, \quad \forall i \in \mathbb{I}, j \in\mathbb{J}.
	\end{cases}
\end{equation}
\par
In conventional NOMA, user pairing is performed either by pairing a strong channel gain near user with a best channel gain far user \cite{16} or by pairing a strong channel gain near user with a weakest channel gain far user \cite{18} for sum rate maximization. In this paper, these two different pairing techniques presented in \cite{16} and \cite{18} are named as greedy pairing and non-greedy pairing respectively, based on their best and worst resource capturing behavior. Additionally, these pairing techniques are used as benchmark for the performance comparison in section V. Furthermore, due to channel ranking performed in \textbf{Algorithm-1}, it is easy to identify the two sets; strong mobile UAVs represented by set $\mathbb{I}$, weak mobile UAVs represented by set $\mathbb{J}$ to define a two-sided matching problem for energy efficient pairing of the mobile UAVs in uplink NOMA. The optimization problem is given as 

\begin{subequations} \label{problem1}
\begin{alignat}{2}
(\textbf{P2}):& \max\limits_{H,P,T_T}
& & (\eta_{EE}) \label{eq:23a}\\ 
& ~ \text{s.~t.~}& & \sum_{k=1}^{K_T} \kappa_k(t) \leq K_T,~~k=\{1,\dots,K_T\}  \label{eq:23b} \\
& & & \sum_{i=1}^{I} \omega(i,j)(t) \leq 1, ~~\forall j \in \mathbb{J} \label{eq:23c}\\
& & & \sum_{j=1}^{J} \omega(i,j)(t) \leq 1, ~~\forall i \in \mathbb{I} \label{eq:23d}\\
& & & M=\sum_{i=1}^{I}\sum_{j=1}^{J}\omega,~~\forall i \in \mathbb{I},\forall j \in \mathbb{J} \label{eq:23e}\\
& & &  P_i-P_j \geq P_{th}. \label{eq:23f}
\end{alignat}
\end{subequations}

\begin{algorithm}[t!]
	\small
	\caption{Solution to Problem (P2)}
	
		\hspace*{\algorithmicindent} \textbf{Initialize:} $\mathbb{W}$,~$UL_{matched,i,j}$,~$UL_{notmatched,i,j}$
	\begin{algorithmic}[1]
		\State \textbf{Generate} $PR_i$,$PR_j$, $\triangleright~ \footnotesize \text{Preference lists of}~$$i \in \mathbb{I}, j \in \mathbb{J} $
		\For {each time instant $t$}
		\For {each $UAV_i$ from $PR_i$}
		\State $UL_{notmatched,i} \leftarrow UAV_{i}$
		\Repeat
		\For {each $UAV_j$ from $PR_j$}  $\triangleright~ \footnotesize|P_i|^2-|P_{j}|^2 \geq P_{th}$
		\State {$UL_{notmatched,j} \leftarrow UAV_{j}$}
		\If {$UAV_{i} \in PR_j$ $\vee$ $UAV_{j} \notin$ $UL_{matched,j}$}
		\If {$UAV_{j} \in Temp(i,j)$ $\wedge$ $E_{i,j}$~\textless~$E_{i,j-1}$ }
		\State $Remove~UAV_j $ from $UL_{notmatched,j}$
		\State $Update~UAV_j $ in $UL_{matched,j}$	
		\State $Remove~UAV_{j-1} $ from $UL_{matched,j}$		
		\Else
		\State $Set~Temp(i,j)\leftarrow~UAV_{i},UAV_{j}$
		\State $\omega(i,j) = 1$
		\EndIf
		\Else
		\State $\omega_(i,j) = 0$		
		\EndIf

		\State $\mathbb{W}(i,j)=\omega(i,j)$
		\EndFor
		\State $Remove~UAV_i $ from $UL_{notmatched,i}$ 
		\State $Update~UAV_i$ in $UL_{matched,i}$
		\Until $UL_{notmatched,i}=\phi,~$Eq.(\ref{eq:23e})
		\EndFor
		\EndFor
	\end{algorithmic}
\end{algorithm}

In the above optimization problem Eq. (\ref{eq:23a}) is given to maximize the energy efficiency of UAVs where the constraint (\ref{eq:23b}) represents the maximum successful uplink transmissions of mobile UAVs if matched under the constraints (\ref{eq:23c}) and (\ref{eq:23d}). Also, the number of matched pairs should be equal the available sub-channels expressed in constraint (\ref{eq:23e}). Moreover, the pairing should be under the power difference constraint (\ref{eq:23f}) required for successful data recovery at the BS in NOMA.
\par
For the above problem (\textbf{P2}), a two-sided one-to-one matching \cite{35,36,37} based solution is proposed to pair the strong and weak mobile UAVs where preference lists of both sides of the mobile UAVs are generated at the BS based on their energy consumption ($E_T$) given in Eq. (\ref{eq066}) to minimize the energy and their power difference threshold to successfully recover the data at the BS. Moreover, two additional lists of mobile UAVs are required; list of matched UAVs i.e., $UL_{matched}$ and list of not matched UAVs i.e., $UL_{notmatched}$. Then, following the matching based (\textbf{Algorithm-2}), a strong mobile UAV from the set of proposers $\mathbb{I}$ is selected to propose the weak mobile UAVs in the set of selectors $\mathbb{J}$ according to the available preference lists represented as

\begin{equation}
	\begin{split}
	\label{eq24}
	PR_{i}(t)=\{j \in J \mid |h_i|^2-|h_{j}|^2  \geq Ch_{th} ~\wedge \\ E_{T,j}<E_{T,j+1},\dots,E_{T,\bar{J}-1}<E_{T,\bar{J}}\},
	\end{split}
\end{equation}

where $PR_{i}$ has the indexes of the $\bar{J}$($\bar{J}\leq J$) selectors such that the selectors at the top of the list is the one that satisfies the channel threshold condition and the minimum energy consumption constraint of the proposer. Thus, the limited number of selectors $\bar{J}$($\bar{J}\leq J$) added to the $PR_{i}$ helps to reduce the paring complexity. Particularly, the goal of the strong mobile UAVs as proposers is to minimize the energy consumption by searching weak mobile UAVs that satisfies their rate at minimum power. While on the other hand, weak mobile UAVs also wants to maintain the power difference threshold condition that is required for successful data recovery at the BS \cite{38,39}. Thus, for the stability of matching, weak mobile UAVs are considered as proposers in the second round to select strong mobile UAVs from the set of selectors. Similar to the (\textbf{Algorithm-2}), weak mobile UAVs from the set of proposers $\mathbb{J}$ are selected one by one to propose the strong mobile UAVs present in their corresponding preference lists from the set of selectors $\mathbb{I}$. The preference list of a weak mobile $UAV_j$ is represented as

\begin{equation}
\small
		\label{eq25}
		PR_{j}(t)=\{i \in I \mid |h_i|^2-|h_{j}|^2  \geq Ch_{th} ~\wedge
		P_i-P_j  \geq P_{th}\}.
\end{equation}
\par

Note that only the difference in the second round is of the preference list $PR_j$ generated for the weak mobile UAVs while the remaining scheme remains the same as the \textbf{Algorithm-2}.
\par
The proposed matching based pairing quickly reaches to the stability state because unlike the distributive approach in \cite{34}, the negotiation process between proposers and selectors for a suitable power allocation is eliminated. The power allocation is performed separately as \textbf{Algorithm-1} to avoid the repeatedly proposals and rejections between the mobile UAVs that creates an extra overhead at the BS. Furthermore, at the end of the \textbf{Algorithm-2} two-sided stability is achieved by matching all of the mobile UAVs until not a single blocking pair $Temp(i,j)$ is left \cite{35}. However, it cannot reach the stability until the list of unmatched pairs becomes empty i.e. $UL_{notmatched}=\phi$ and the number of matched pairs equals to the number of available subchannels.


\section{Performance Analysis and Pairing Complexity}
In this section, comparison and analysis of the proposed pairing technique is performed for mobile UAVs considering the static ground users as already paired users. In order to analyze the performance of the proposed pairing solution for mobile UAVs according to \textit{Remark-1}, a brief analysis for mobile UAVs is performed and for that a simple optimization problem is defined to compare the performance gains in terms of data rate satisfaction of the mobile UAVs at the BS. As it is mentioned earlier in \textbf{Definition-3} that $\kappa$ represents the number of mobile UAVs whose target rates at the BS are successfully achieved. Therefore, the maximum sum of $\kappa(t)$ represents the highest number of mobile UAVs with successful uplink transmission at time ($t$). Thus, problem \textbf{(P3)} can be defined as

\begin{subequations} \label{problem3}
	\begin{alignat}{2}
		(\textbf{P3}):& \max\limits_{\kappa \in \{0,1\}}
		& & \sum_{k=1}^{K_T} \kappa_k(t), \label{eq:26a}\\ 
		& ~ \text{s.~t.~}& & R_k \geq R_k^T, \quad \forall~k=\{1,2,\dots,K_T\}. \label{eq:26b}
	\end{alignat}
\end{subequations}
\par

The above optimization problem \textbf{(P3)} is maximized if and only if \\

\begin{equation}
	\label{eq20}
\sum_{k=1}^{K_T} \kappa_k(t) \leq K_T,
\end{equation}

\begin{figure}[t!]
\centering
\includegraphics[scale=0.6]{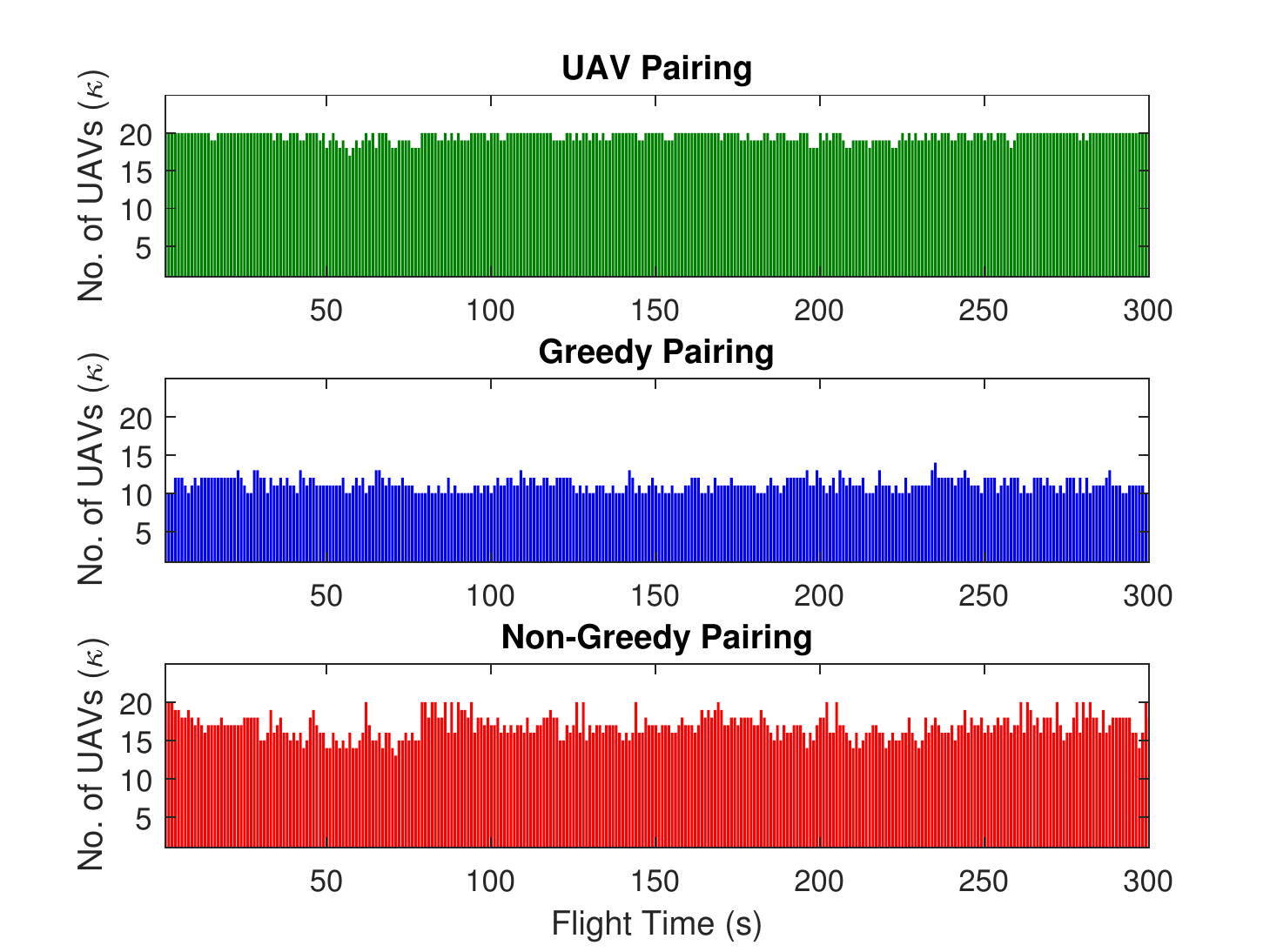}
\caption{Number of UAVs ($\kappa$) from total number of UAVs ($K_T$) whose target rate is successfully achieved at the BS with minimized transmit powers.}
\label{Figure:33}
\end{figure}

Applying the power optimization (\textbf{Algorithm-1}) and UAV pairing (\textbf{Algorithm-2}) and a  simulation is performed with $K_T=20$ mobile UAVs. The number of mobile UAVs that fulfills the target rate constraint (\ref{eq:26b}) at each time instant $t$ is calculated and graphically represented as Fig. \ref{Figure:33}.
\par
In Fig. \ref{Figure:33}, the target rate satisfaction of the proposed UAV pairing technique is higher than the greedy and non-greedy pairing because both of the later techniques are meant to maximize the sum rate without considering the individual target rates of the mobile UAVs. Therefore, in greedy pairing technique the strong mobile UAVs after greedily capturing the best mobile UAVs among weak UAVs do not take care of their required rate and fails to satisfy the target rates of almost 50\% of the mobile UAVs in the network. While in non-greedy pairing technique without considering the channel difference threshold the mobile UAVs with high channel difference and non-optimized powers are paired together and as a result large number of mobile UAVs fails to achieve their required target rates at the BS. Thus, resulting as an unbalanced pairing of the mobile UAVs. Moreover, the target rate satisfaction of the proposed technique is not 100\% because we keep the 10\% tolerance for mobile UAVs whose target rates are not fulfilled and the pairing/unpairing is performed only when the number of UAVs with unsatisfied target rate crosses that tolerance level. Therefore, a balanced UAV pairing technique is performed to satisfy their target rates with minimum energy consumption and low pairing complexity.

\par
Reducing the pairing complexity of the mobile UAVs in a NOMA UAV networks with large number of aerial users is an open research challenge and requires an optimal pairing for the mobile UAVs that continuously change their positions during mobility. Therefore, it is not claimed in this paper that the proposed pairing technique has low complexity at the BS but it is shown that the complexity is reduced by limiting the pairing/unpairing of mobile UAVs based on their channel and power difference threshold constraints. At least, the proposed UAV pairing reduces the pairing overhead at the BS.

\section{Simulation Results and Discussion}
In this section, simulation results of the proposed UAV pairing and power allocation technique are provided to evaluate the performance compared to the conventional fixed pairing techniques in NOMA. For simulations, normalized values of the bandwidth $B$=1Hz, $P_c$=1W, distance $d$=0.1-1 are considered while the height $H$ and flight energy consumption $e_f$ of the mobile UAVs are assumed as constant values. The probabilistic channel model based on the angle of elevation ($\theta$) of mobile UAVs is used for all simulation results where LOS($Pr^{LOS}$) and NLOS($Pr^{NLOS}$) probabilities are used to calculate the combined path loss of ($PL$) of mobile UAVs as shown in the Fig. \ref{fig:figure3}. Additionally, for probabilistic LOS channel model in dense urban environment variables similar to \cite{21} are set as  $\zeta=12.0870$, $\delta=0.1139$, and additional loss factors due to A2G links are set as $\Lambda_{LOS}$=1.6, $\Lambda_{NLOS}$=23, with path loss exponent $\psi$=2. Furthermore, for the mobility of the mobile UAVs RWP mobilit model is used where \{$x,y$\} position and speed intervals are also normalized that depicts the mobility of an average speed UAV with normalized distance variations.
\par
\begin{figure}[t!]
	\centering
	\null\hfill
	\subfloat[Energy efficiency of the mobile UAVs throughout their flight duration ($T_T$=300s).] {\includegraphics[width=0.5\textwidth]{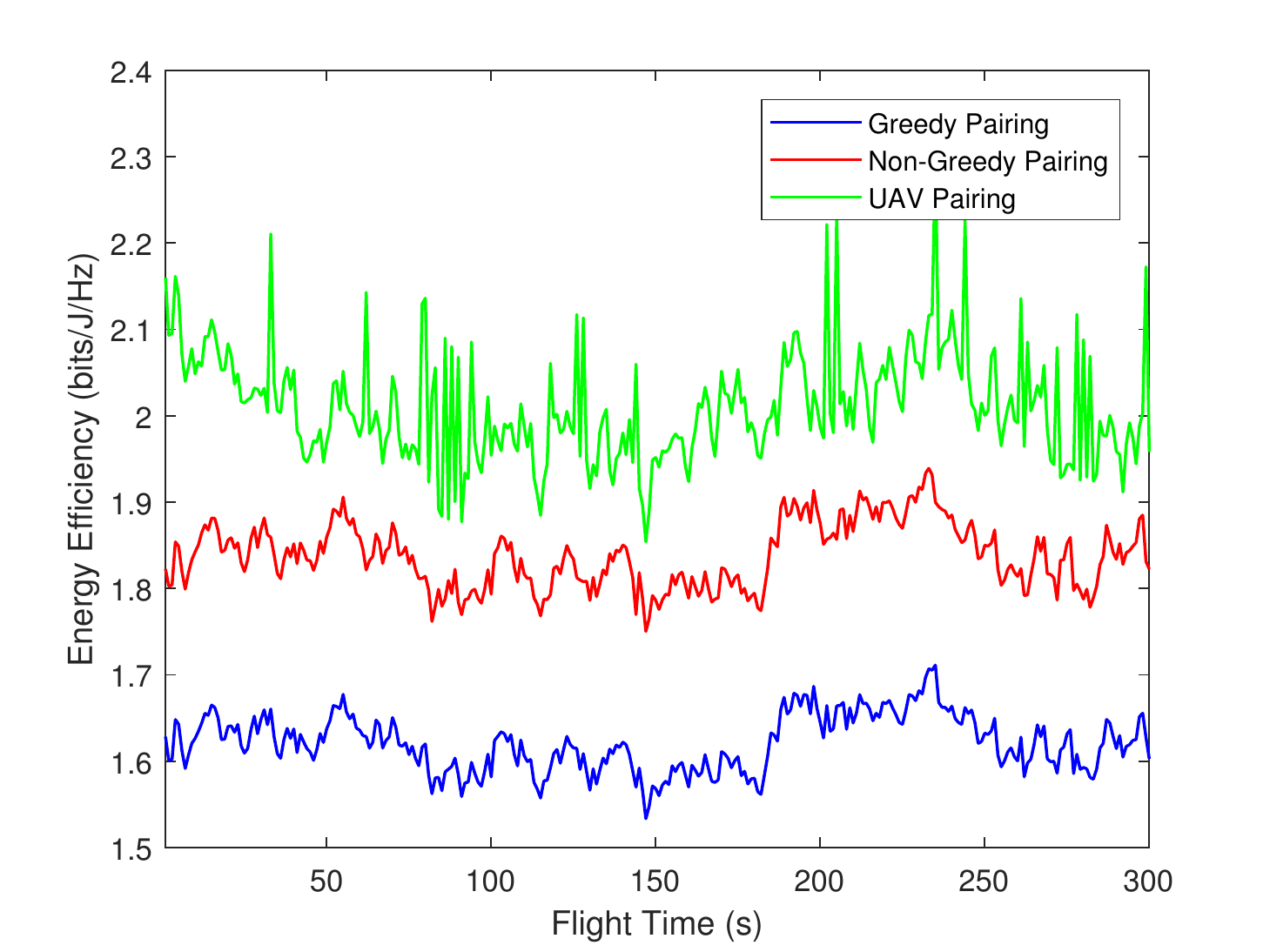}}
	\hfill
	\subfloat[Energy consumption of the mobile UAVs in Joules.]{\includegraphics[width=0.5\textwidth]{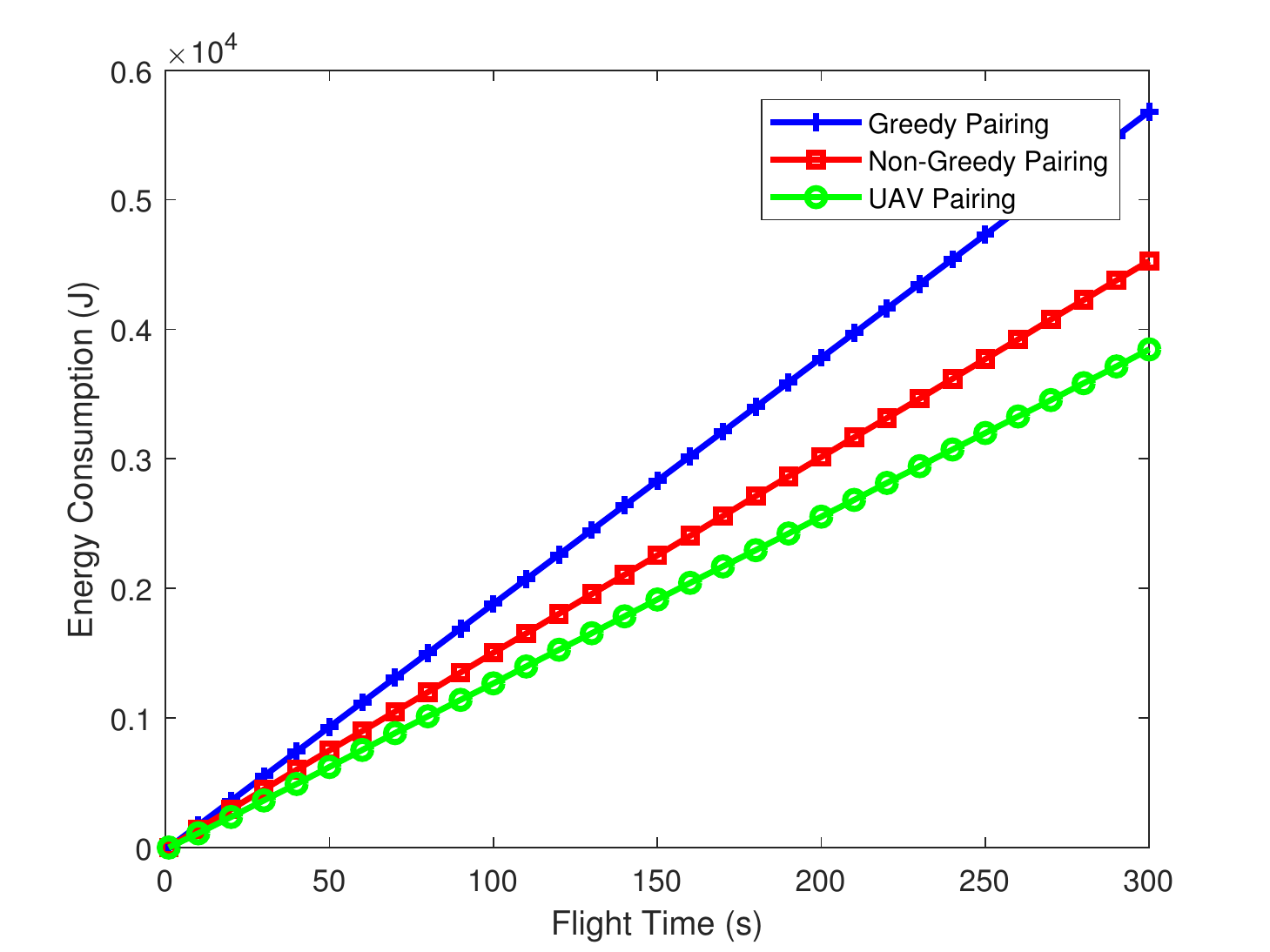}}
	\hfill\null
	\caption{\label{Figure:4}
		Comparison of the proposed UAV pairing with conventional NOMA pairing techniques in terms of energy efficiency while considering $K_T$=20 mobile UAVs for the total flight duration of $T_T$=300s.}
\end{figure}
\begin{figure}
	\centering
	\includegraphics[scale=0.6]{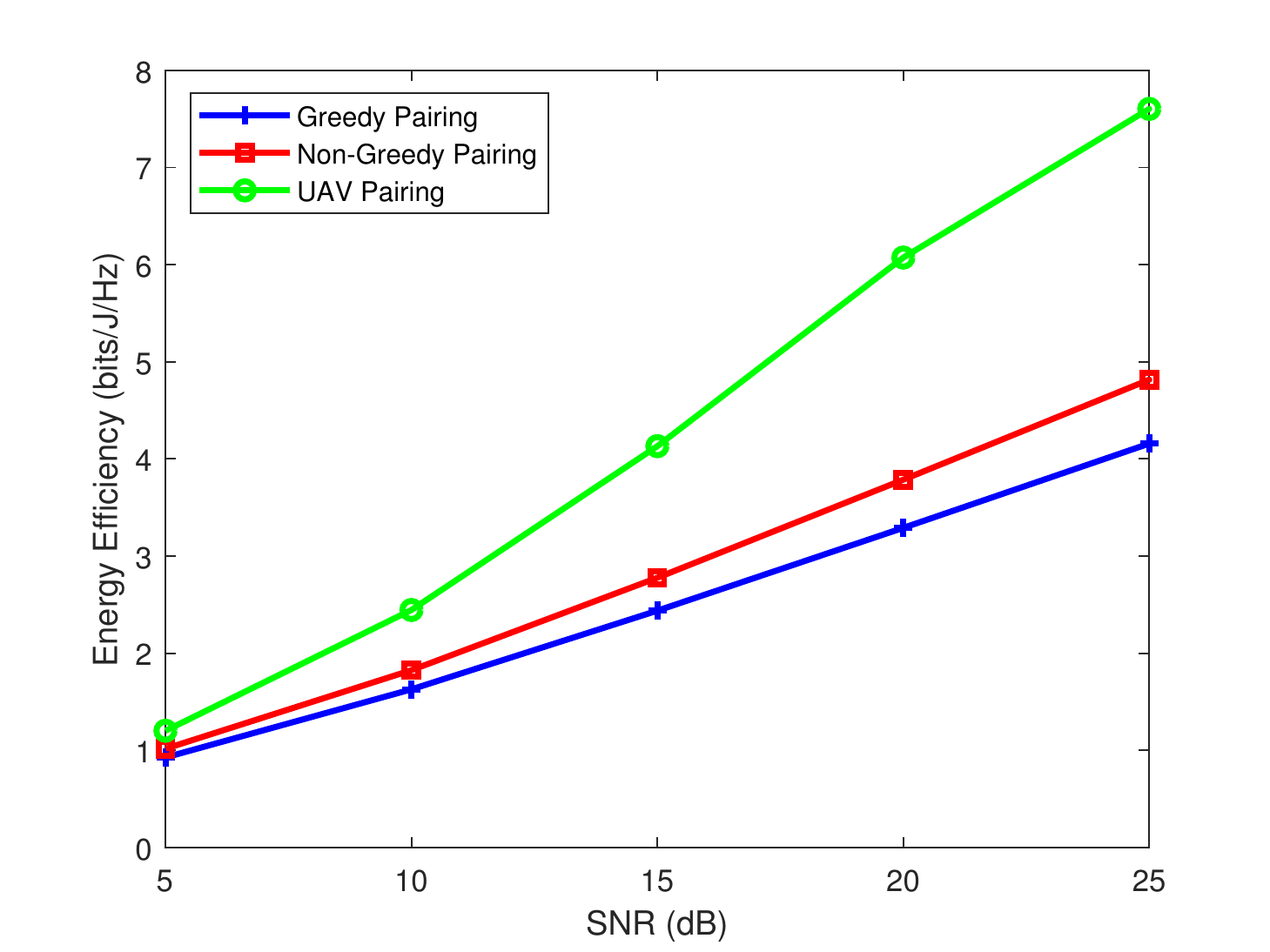}
	\caption{Energy efficiency analysis at different SNR values with $K_T$=20 mobile UAVs.}
	\label{Figure:5}
\end{figure}

Initially, the simulations are performed for the flight time of 300 seconds while considering the above parameters at SNR=10dB, the results are shown in Fig.\ref{Figure:4}. Then, for the performance at higher SNR and with large number of users $K_T$ results are shown in Fig. \ref{Figure:5}-\ref{Figure:7}. In Fig. \ref{Figure:4}(a), energy efficiency of $K_T$=20 mobile UAVs is shown at each time instant of mobile UAVs total flight duration of 300s. The performance of the proposed UAV pairing is better than the conventional NOMA pairing techniques throughout the flight duration due to the power optimization performed before pairing. Moreover, matching based pairing based on the minimum energy of the mobile UAVs further minimizes their energy consumption and increases the energy efficiency. It is noticed in Fig. \ref{Figure:4}(b) that the energy consumption of the greedy pairing is higher than the other two pairing techniques because it greedily captures the best mobile UAVs that may maximize the sum rate using higher transmit powers but minimizes the energy efficiency of the mobile UAVs with limited battery power. On the other hand, the energy efficiency of non-greedy pairing technique is better than the greedy pairing due to using the fractional powers calculated for each mobile UAV based on their channel gain. But, pairing the mobile UAVs based on a high channel difference i.e. pairing a strong channel UAV with weakest channel gain and less transmit powers fails fulfill the target rates with these minimum powers and makes an unbalanced pairing of mobile UAVs. Hence, it is clear from the simulation results in Fig. \ref{Figure:33} and Fig. \ref{Figure:4} that conventional fixed pairing techniques fails to fulfill the individual target rates of the mobile UAVs despite of consuming high energy in a cellular-connected UAV network.
\par

\par
Furthermore, the energy efficiency comparison of the proposed technique at low and high SNR is performed in Fig.\ref{Figure:5}. High energy efficiency of the proposed UAV pairing technique with increasing difference shows its performance gain compared to the conventional fixed pairing techniques. Moreover, the performance comparison for large number of aerial users shows the performance gain of the proposed technique.

\section{Research Challenges and Recommendations}
In this section, the potential research challenges of the cellular-connected UAV networks with largely deployed mobile aerial users are discussed. Unlike the cellular ground users the cellular-connected aerial users have different characteristics due to which, conventional cellular channel models can not be used for the UAV communications especially when UAVs are mobile. Therefore, many of the research issues in cellular-connected mobile UAV communications remain open and outlined as follows:
\par
Most prior and important issue of largely deployed cellular-connected UAVs is the spectrum and resource sharing with the cellular users. This problem is similar to the resource sharing problem in cognitive radio networks where the resources of the primary users are shared with the cognitive users while guaranteeing the QoS requirements of the primary users. In cellular-connected UAVs, spectrum and resources of the cellular users on ground need to be efficiently shared as the priority of cellular users at the BS is higher than the UAVs. One of the solution is to share the spectrum and resources among both types of users on their need and QoS requirements. Although the needs and fulfillment of QoS requirements of a cellular user on ground are more important than a cellular-connected UAV.  
\par
Mobility of the UAVs is another important issue for aerial users. It can be divided into two types: (1) High mobility profile UAVs that mainly includes high speed fixed-wing UAVs, (2) Normal or average mobility profile UAVs that includes rotatory wing-UAVs with average speed, and (3) Low mobility profile UAVs are rotatory-wing UAVs that moves very less only to reach at certain destination point and starts hovering over there for data collection or to provide some services. Channel model and transmission requirements of all these above three types of UAVs are different and need to be well-designed. Especially in high and average speed UAVs the channel condition continuously varies that affects the transmission. Moreover, it is more challenging in a heterogeneous UAV network to handle the UAVs with three different types of mobility profiles. Also, the channel modeling of mobile UAVs depends on the environment for which it is developed. For example, the channel statistics of high and average speed UAVs in urban area will be different from the dense urban environments due to direct LOS and NLOS links between the UAVs and the GBS.
\par
Interference management and mitigation for a reliable transmission is another problem especially if largely deployed UAVs transmit simultaneously to a single BS as the only centralized receiving entity. It can be handled in a distributed manner that can reduce the transmission overhead. Additionally, uplink transmission of largely deployed UAVs in a grant-free manner can be a potential research direction. Moreover, massive multiple input multiple output (MIMO) with very large number of antennas with beamforming technique for interference issue and reconfigurable intelligent surfaces (RIS) for reliable transmission can be useful in cellular-connected heterogeneous UAV networks.

\section{Conclusions}
In this paper, an energy efficient NOMA based uplink transmission scheme was presented for cellular-connected UAVs under mobility constraint where channel conditions of mobile UAVs are continuously changing. Particularly, there were three main objectives of this work; (i) to investigate the channel characteristics of mobile UAVs that were continuously changing with their speeds and positions (ii) to minimize the energy consumption of mobile UAVs during uplink data transmission, (ii) energy efficient user pairing of mobile UAVs to guarantee their QoS rate requirements with less energy. Therefore, a probabilistic channel model for these largely deployed mobile aerial users considering their LOS and NLOS link probabilities in a dense urban region was considered. The problem was efficiently solved by applying a fast convergence power optimization algorithm to minimize their transmission powers and guaranteeing their asymmetric QoS requirements. Then, a matching based pairing technique was used for energy efficiency. Finally, pairing complexity and performance of the proposed technique was analyzed to show its effectiveness and performance gain compared to the conventional NOMA technique in terms of energy efficiency and target rate achievement.
\ifCLASSOPTIONcaptionsoff
\fi


\begin{thebibliography}{1}

\bibitem{1}
Gupta L, Jain R, Vaszkun G. \emph{"Survey of important issues in UAV communication networks."} IEEE Communications Surveys and Tutorials. 2015 Nov 3;18(2):1123-52.

\bibitem{2}
Zeng Y, Wu Q, Zhang R. \emph{"Accessing from the sky: A tutorial on UAV communications for 5G and beyond."} Proceedings of the IEEE. 2019 Dec 2;107(12):2327-75.

\bibitem{3}
S. D. Muruganathan, X. Lin, H.-L. Maattanen, Z. Zou, W. A. Hapsari, and S. Yasukawa, “An overview of 3GPP release-15 study on enhanced LTE support for connected drones,” May 2018. [Online]. Available: https://arxiv.org/ftp/arxiv/papers/1805/1805.00826.pdf

\bibitem{4}
B. Van der Bergh, A. Chiumento, and S. Pollin, “LTE in the sky: Trading off propagation benefits with interference costs for aerial nodes,” IEEE Commun. Mag., vol. 54, no. 5, pp. 44–50, May 2016.

\bibitem{5}
Qualcomm Technologies, Inc., “LTE unmanned aircraft systems trial report,” San Diego, CA, USA, Rep., May 2017. [Online]. Available:
https://www.qualcomm.com/documents/lte-unmanned-aircraft-systemstrial-report

\bibitem{6}
Liu Y, Qin Z, Cai Y, Gao Y, Li GY, Nallanathan A. \emph{"UAV communications based on non-orthogonal multiple access."} IEEE Wireless Communications. 2019 Feb 13;26(1):52-7.

\bibitem{7}
Wang W, Tang J, Zhao N, Liu X, Zhang XY, Chen Y, Qian Y. \emph{"Joint Precoding Optimization for Secure SWIPT in UAV-Aided NOMA Networks."} IEEE Transactions on Communications. 2020 Apr 28.

\bibitem{8}
Liu X, Wang J, Zhao N, Chen Y, Zhang S, Ding Z, Yu FR. \emph{"Placement and power allocation for NOMA-UAV networks."} IEEE Wireless Communications Letters. 2019 Mar 8;8(3):965-8.

\bibitem{9}
Federal Aviation Administration, USA, “FAA Aerospace Forecast Fiscal Years 2020-2040 Full Forecast Document and Tables, USA, Rep., 2020. [Online]. Available:
https://www.faa.gov/data$\_$research/aviation/aerospace$\_$forecasts/media/ \\ Unmanned$\_$Aircraft$\_$Systems.pdf

\bibitem{10}
Zhan C, Zeng Y. \emph{"Energy-Efficient Data Uploading for Cellular-Connected UAV Systems."} IEEE Transactions on Wireless Communications. 2020 Jul 27.

\bibitem{11}
Zeng Y, Lyu J, Zhang R. \emph{"Cellular-connected UAV: Potential, challenges, and promising technologies."} IEEE Wireless Communications. 2018 Sep 24;26(1):120-7.

\bibitem{12}
Mozaffari M, Saad W, Bennis M, Nam YH, Debbah M. A tutorial on UAVs for wireless networks: Applications, challenges, and open problems. IEEE Communications Surveys and Tutorials. 2019 Mar 5;21(3):2334-60.

\bibitem{13}
R16-38812 Y. \emph{"Technical Specification Group Radio Access Network: Study on Non-Orthogonal Multiple Access (NOMA) for NR: 3GPP Technical Report"} (2018).

\bibitem{14}
David K, Berndt H. \emph{"6G vision and requirements: Is there any need for beyond 5G?. IEEE Vehicular Technology Magazine."} 2018 Jul 18;13(3):72-80.

\bibitem{15}
Latva-aho, Matti, Kari Leppänen, Federico Clazzer, and Andrea Munari. \emph{"Key drivers and research challenges for 6G ubiquitous wireless intelligence."} (2020).

\bibitem{15a}
Ye, Neng, et al, \emph{"Uplink nonorthogonal multiple access technologies toward 5G: A survey."} Wireless Communications and Mobile Computing 2018 (2018).

\bibitem{15b}
Mohammadkarimi, Mostafa, Muhammad Ahmad Raza, and Octavia A. Dobre, \emph{"Signature-based nonorthogonal massive multiple access for future wireless networks: Uplink massive connectivity for machine-type communications."} IEEE Vehicular Technology Magazine 13.4 (2018): pp. 40-50.

\bibitem{15c}
M. B. Shahab, R. Abbas, M. Shirvanimoghaddam and S. J. Johnson, \emph{"Grant-Free Non-Orthogonal Multiple Access for IoT: A Survey,"} in IEEE Communications Surveys \& Tutorials, vol. 22, no. 3, pp. 1805-1838, third quarter 2020, doi: 10.1109/COMST.2020.2996032.

\bibitem{15d}
Y. Saito, Y. Kishiyama, A. Benjebbour, T. Nakamura, A. Li and K. Higuchi, \emph{"Non-Orthogonal Multiple Access (NOMA) for Cellular Future Radio Access,"} 2013 IEEE 77th Vehicular Technology Conference (VTC Spring), Dresden, 2013, pp. 1-5.

\bibitem{16}
Ding Z, Fan P, Poor HV. Impact of user pairing on 5G nonorthogonal multiple-access downlink transmissions. IEEE Transactions on Vehicular Technology. 2015 Sep 22;65(8):6010-23.

\bibitem{17}
Shahab MB, Kader MF, Shin SY. A virtual user pairing scheme to optimally utilize the spectrum of unpaired users in non-orthogonal multiple access. IEEE Signal Processing Letters. 2016 Oct 19;23(12):1766-70.

\bibitem{18}
Shahab MB, Shin SY. User pairing and power allocation for non-orthogonal multiple access: Capacity maximization under data reliability constraints. Physical Communication. 2018 Oct 1;30:132-44.

\bibitem{19}
Sedaghat MA, Müller RR. On user pairing in uplink NOMA. IEEE Transactions on Wireless Communications. 2018 Mar 20;17(5):3474-86.

\bibitem{20}
Azam I, Shahab MB, Shin SY. Role switching and power allocation technique for mobile users in non-orthogonal multiple access. Physical Communication. 2020 Aug 13:101179.

\bibitem{21}
Sohail MF, Leow CY, Won S. \emph{"Energy-Efficient Non-Orthogonal Multiple Access for UAV Communication System."} IEEE Transactions on Vehicular Technology. 2019 Sep 3;68(11):10834-45.

\bibitem{22}
Zeng Y, Zhang R. \emph{"Energy-efficient UAV communication with trajectory optimization."} IEEE Transactions on Wireless Communications. 2017 Mar 28;16(6):3747-60.

\bibitem{23}
F. Cui, Y. Cai, Z. Qin, M. Zhao, and G. Y. Li, “Multiple access for mobile-UAV enabled networks: Joint trajectory design and resource allocation,” IEEE Trans. Commun., vol. 67, no. 7, pp. 4980–4994, Jul. 2019.

\bibitem{24}
Mu X, Liu Y, Guo L, Lin J. Non-Orthogonal Multiple Access for Air-to-Ground Communication. IEEE Transactions on Communications. 2020 Feb 11;68(5):2934-49.

\bibitem{25}
Mei W, Zhang R. Uplink cooperative NOMA for cellular-connected UAV. IEEE Journal of Selected Topics in Signal Processing. 2019 Feb 13;13(3):644-56.

\bibitem{26}
Liu M, Gui G, Zhao N, Sun J, Gacanin H, Sari H. UAV-Aided Air-to-Ground Cooperative Non-orthogonal Multiple Access. IEEE Internet of Things Journal. 2019 Dec 3;7(4):2704-15.

\bibitem{27}
Althunibat S, Badarneh OS, Mesleh R. \emph{"Random waypoint mobility model in space modulation systems."} IEEE Communications Letters. 2019 Mar 27;23(5):884-7.

\bibitem{27a}
U. Mengali and A. N. D’Andréa, Synchronization Techniques for Digital Receivers. New York, NY, USA: Springer, 1997.

\bibitem{28}
Enhanced LTE Support for Aerial Vehicles, document 3GPP-TR-36.777, 3GPP, 2017. [Online]. Available: https://portal.3gpp.org/desktopmodules/Specifications/SpecificationDetails\\.aspx?specificationId=3231

\bibitem{29}
Liu Y, Qin Z, Cai Y, Gao Y, Li GY, Nallanathan A. UAV communications based on non-orthogonal multiple access. IEEE Wireless Communications. 2019 Feb 13;26(1):52-7.

\bibitem{30}
A. Al-Hourani, S. Kandeepan, and S. Lardner, “Optimal LAP altitude for maximum coverage,” IEEE Wireless Commun. Lett., vol. 3, no. 6, pp. 569–572, Dec. 2014.

\bibitem{31}
Khuwaja AA, Chen Y, Zhao N, Alouini MS, Dobbins P. A survey of channel modeling for UAV communications. IEEE Communications Surveys and Tutorials. 2018 Jul 16;20(4):2804-21.

\bibitem{32}
A. Al-Hourani, S. Kandeepan, and A. Jamalipour, “Modeling air-toground path loss for low altitude platforms in urban environments,” in Proc. IEEE Global Commun. Conf., Dec. 2014, pp. 2898–2904.

\bibitem{33}
A. Filippone, “Flight performance of fixed and rotary wing aircraft”. Washington, DC, USA: AIAA, 2006

\bibitem{34}
Liang W, Ding Z, Li Y, Song L. User pairing for downlink non-orthogonal multiple access networks using matching algorithm. IEEE Transactions on communications. 2017 Aug 25;65(12):5319-32.

\bibitem{35}
Gu Y, Saad W, Bennis M, Debbah M, Han Z. Matching theory for future wireless networks: Fundamentals and applications. IEEE Communications Magazine. 2015 May 14;53(5):52-9.

\bibitem{35a}
Sun Q, Han S, Chin-Lin I, Pan Z. On the ergodic capacity of MIMO NOMA systems. IEEE Wireless Communications Letters. 2015 Apr 27;4(4):405-8.

\bibitem{36}
D. Gusfield and R. W. Irving, The Stable Marriage Problem: Structure and Algorithms. Cambridge, MA, USA: MIT Press, 1989.

\bibitem{37}
A. Roth and M. Sotomanyor, Two Sided Matching: A Study in Game-Theoretic Modeling and Analysis, 1st ed. Cambridge, U.K.: Cambridge Univ. Press, 1989.

\bibitem{38}
Shahab MB, Kader MF, Shin SY. On the power allocation of non-orthogonal multiple access for 5G wireless networks. In2016 international conference on open source systems and technologies (ICOSST) 2016 Dec 15 (pp. 89-94). IEEE.

\bibitem{39}
Chen X, Zhang Z, Zhong C, Ng DW. Exploiting multiple-antenna techniques for non-orthogonal multiple access. IEEE Journal on Selected Areas in Communications. 2017 Jul 7;35(10):2207-20.


\end{thebibliography}
\end{document}